# Nanoscale rectification at the LaAlO$_3$/SrTiO$_3$ interface

Daniela F. Bogorin[1], Chung Wung Bark[2], Ho Won Jang[2], Cheng Cen[1], Chad M. Folkman[2], Chang-Beom Eom[2], Jeremy Levy[1]

[1]Department of Physics and Astronomy,

University of Pittsburgh, Pittsburgh, Pennsylvania 15260, USA

[2]Department of Materials Science and Engineering,

University of Wisconsin-Madison, Madison, Wisconsin 53706, USA

jlevy@pitt.edu

ABSTRACT

Control over electron transport at scales that are comparable to the Fermi wavelength or mean-free path can lead to new families of electronic devices. Here we report electrical rectification in nanowires formed by nanoscale control of the metal-insulator transition at the interface between LaAlO$_3$ and SrTiO$_3$. Controlled in-plane asymmetry in the confinement potential produces electrical rectification in the nanowire, analogous to what occurs naturally for Schottky diodes or by design in structures with engineered structural inversion asymmetry. Nanostructures produced in this manner may be useful in developing a variety of nanoelectronic, electro-optic and spintronic devices.

The discovery of a high-mobility two-dimensional electron gas (2DEG) at the interface between LaAlO$_3$ and SrTiO$_3$[1,2] has opened exciting new opportunities for electric-field controlled phenomena[3-10] and devices[11,12]. The interface between thin films of LaAlO$_3$ and TiO$_2$-terminated SrTiO$_3$ exhibits an abrupt insulator-to-metal transition with increasing LaAlO$_3$ thickness. The 2DEG is *n*-type and strongly localized at the interface[13]. Films grown at a critical thickness of 3 unit cells (3uc-LAO/STO) exhibit a reversible and hysteretic interfacial metal-insulator phase transition that can be programmed by voltages $V_{bg}$~+/-100 V applied to the back SrTiO$_3$ substrate[3].

Nanoscale control of the interfacial metal-insulator transition in 3uc-LAO/STO can be achieved using a conducting AFM probe[11]. A positive voltage applied to the AFM tip with respect to the interface locally switches the interface into a conducting state, while a negative voltage locally restores the interface to an insulating state. Conducting nanostructures are created by scanning the AFM tip over the 3uc-LAO/STO surface along a trajectory $(x(t),y(t))$ while a voltage $V_{tip}(t)$ is applied (Fig. 1). Writing with a smaller $V_{tip}$ or faster scan rate generally produces a narrower, less highly conducting nanostructure. Nanostructures are stable for ~1 day in atmospheric conditions at room temperature, and indefinitely under modest vacuum[12]. It is believed that the AFM writing procedure charges the top LaAlO$_3$ surface and modulation-dopes the interface with near-atomic spatial precision. Using this writing procedure, a variety of quasi-zero-dimensional and quasi-one-dimensional nanostructures have been created[11]. Nanoscale tunnel junctions and transistors with features as small as 2 nm have also been demonstrated[12].

Asymmetries in the electronic confining profile generally lead to non-reciprocal behavior in transport (i.e. $I(V) \neq -I(-V)$). Diodes formed by *p-n* junctions, modulation-doped heterostructures, and Schottky barriers are just a few examples of non-reciprocal devices that play an essential role in modern electronics . Non-reciprocal devices down to nanoscale dimensions have been created by various methods including metal-semiconductor[14] and semimetal-semiconductor interfaces[15], controlled in-plane doping[16], complementary-doped nanotubes[17], and hybrid organic/inorganic semiconductors[18].

In this letter, we report the creation of nanoscale rectifying junctions along nanowires formed within 3uc-LAO/STO heterostructures. Thin films of $LaAlO_3$ are grown on $TiO_2$-terminated (001) $SrTiO_3$ substrates[19] by pulsed laser deposition with *in situ* high pressure reflection high energy electron diffraction (RHEED)[20]. The films are grown at a substrate temperature of 550C under oxygen pressure of $10^{-3}$ mbar and cooled down to room temperature at a $10^{-3}$ mbar. After growth, electrical contacts to the interface are defined by optical lithography using a combination of ion milling and Au/Ti deposition. Within a 35×35 μm$^2$ "canvas" defined by the electrode edges, nanostructures are "written" and "erased" at the interface using conducting AFM lithography[11, 12] (Fig. 1). The nanoscale writing and subsequent transport experiments are performed at room temperature (295 K) under atmospheric conditions (35-50% relative humidity). During and after the writing process, the transport properties of these nanostructures are monitored by applying a small voltage $V_s$ =0.1V to an electrode connected to one end of the nanowire. The resulting current is measured at a second electrode connected to the other end of the nanowire which is held at virtual ground (Fig. 1).

The series of three experiments described below demonstrates how the profile created by the AFM voltage can lead to linear or rectifying behavior. These experiments were performed on the same canvas, whereby a nanowire is first written, followed by transport experiments, followed by an erasure of the canvas (raster-scanning the entire area with $V_{tip}$=-10 V) so that a new nanostructure can be created. Experiments were performed on multiple devices written onto two different 3uc-LAO/STO heterostructures grown under nominally identical conditions. For the purpose of comparison, the results shown here are for devices and structures written in a single experimental session.

In an initial experiment, a nanowire is created with a spatially uniform positive voltage $V_{tip}$=+10 V (Fig. 2(a)). The resulting surface charge is intentionally uniform (Fig. 2(b)) and produces a nanowire along the $x$-direction with highly linear current-voltage ($I$-$V$) characteristics (Fig. 2(d)). The lateral width of this nanowire is determined by cutting it: a negative bias $V_{tip}$=-10 V is applied to the AFM tip as it moves across the wire (along the $y$-direction), and the conductance of the nanowire is monitored using a lock-in amplifier. Analysis of the conductance profile as the nanowire is cut yields a width $w$=2.6 nm.

The nanowire is subsequently erased and replaced by a structure created with an asymmetric sawtooth-shaped voltage pulse (Fig. 2(e)) described by $V_+(x,y)$, which is defined below:

$$V_\pm(x,y)/V_0 \equiv \begin{cases} \pm(2x-1)/x_d, 0 > x < x_d \\ 1, \text{otherwise} \end{cases},$$

where $V_0 = +10$ V and $x_d$=40 nm are the sawtooth amplitude, asymmetry direction and width, respectively. During the writing process, the AFM tip is scanned at a speed $v_x$=400 nm/s. After writing, the measured $I$-$V$ curve (Fig. 2(h)) becomes highly non-reciprocal and rectifying, allowing substantial current flow only for positive bias. There is a small leakage current for the reverse bias, and an onset for reverse-field "breakdown" for $V < -4.5$ V.

The nanostructure is subsequently erased and a third one is written with an asymmetric voltage in the opposite direction $V_-(x,y)$, using parameters $V_0 = +10$ V and $x_d$=40 nm (Fig. 2(i)). Again, the *I-V* curve shows rectifying behavior but with an opposite polarity (Fig. 2(l)). A comparison of the two diode structures (Fig. 2(h,l)) shows a ~20% variation in the reverse breakdown voltage; these variations are typical of those seen experimentally for a given set of parameters.

Schematic energy diagrams for nanostructures written under uniform and non-uniform tip voltage profiles are presented in Fig. 2(c,g,k). The asymmetric conduction-band profile $E_c(x)$ (Fig. 2(g,k)) and *I-V* characteristics mimic that of a metal-semiconductor Schottky junction. Under forward bias (Fig. 2(g)), substantial current flow is observed above a threshold voltage $V>V_{th}$. This threshold voltage allows one to estimate the strength of the electrostatic field produced by the potential gradient $E \approx V_{th}/x_d = 2.5 \times 10^6$ V/cm. Under a reverse bias (Fig. 2(k)), current flow is suppressed by the sharp potential barrier. The rectifying properties of the nanostructures described above depend upon the modulation-doping profile along the nanowire. As demonstrated in Fig. 2, these profiles can be created by spatial modulation of the writing voltage $V_{tip}$. Here we demonstrate a second method for producing non-reciprocal nanostructures. In this approach, spatial variations in the conduction-band profile are created by a precise sequence of erasure steps. In a first experiment, a conducting nanowire is created using $V_{tip}$=+10 V. The initial *I-V* curve (Fig. 3(a), green curve) is highly linear and reciprocal. This nanowire is then cut by scanning the AFM tip across the nanowire at a speed $v_y$=100 nm/s using $V_{tip}$=-2 mV at a fixed location ($x$=20 nm) along the length of the nanowire. This erasure process increases the conduction-band minimum $E_c(x)$ locally by an amount that scales monotonically with the number of passes $N_{cut}$ (Fig. 3(a) inset); the resulting nanostructure exhibits a crossover from conducting to activated to tunneling behavior[12]. Here we focus on the symmetry of the full *I-V* curve. As $N_{cut}$ increases, the transport becomes increasingly nonlinear; however, the *I-V* curve remains highly reciprocal.

The canvas is subsequently erased and a uniform conducting nanowire is written in a similar fashion as before ($V_{tip}$=+10 V, $v_x$=400 nm/s). A similar erasure sequence is performed; however, instead of cutting the nanowire at a single *x* coordinate, a sequence of cuts is performed at nine adjacent *x* coordinates along the nanowire (separated by $\Delta x$=5 nm). The number of cuts at each location along the nanowire $N_{cut}(x)$ increases monotonically with *x*, resulting in a conduction band profile $E_c(x)$ that is asymmetric by design (Fig. 3(b), inset). The resulting *I-V* curve for the nanostructure evolves from being highly linear and reciprocal before writing (Fig. 3(b), green curve) to highly nonlinear and non-reciprocal (Fig. 3(b), red curve).

Nanoscale control over asymmetric potential profiles at the interface between $LaAlO_3$ and $SrTiO_3$ can have many potential applications in nanoelectronics and spintronics. Working as straightforward diodes, these junctions can be used to create half-wave and full-wave rectifiers for AC-DC conversion or for RF detection and conversion to DC. By cascading two or more such junctions, with a third gate for tuning the density in the intermediate regime could form the basis for low-leakage transistor devices. Generally speaking, the ability to control the potential $V(x)$ along a nanowire could be used to create wires with built-in polarizations similar to those created in heterostructures that lack inversion symmetry[21]. Nanoscale control over inversion symmetry breaking could in principle be used to produce nonlinear optical frequency conversion (i.e., second-harmonic generation or difference frequency mixing), thus providing a means for the generation of local sources of light or THz radiation. A straightforward modification of this idea involves the creation of potential profile asymmetries that are in-plane and *transverse* to the nanowire direction (i.e., along the *y* direction in Fig. 1(a)). Such asymmetries could give rise to significant Rashba spin-orbit interactions[22, 23]. The resulting effective magnetic fields could allow control over spin precession along two orthogonal axes[24], and thus exert full three-dimensional control over electron spin[25] in a nanowire.

**ACKNOWLEDGEMENTS**

Support from DARPA seedling (W911N3-09-10258) (J.L.), ARO MURI (W911NF-08-1-0317) (J.L.), and National Science Foundation through grants ECCS-0708759 (C.B.E.), DMR-0704022 (J.L.) and DMR-0906443 (C.B.E.) is gratefully acknowledged.

**FIGURE LEGENDS**

Figure 1. (Color online) Schematic illustrating of the nanowriting process at the LaAlO$_3$/SrTiO$_3$ interface. Au electrodes (shown in yellow) are electrically contacted to the LaAlO$_3$/SrTiO$_3$ interface. The AFM tip with an applied voltage is scanned once between the two electrodes with a voltage applied $V_{tip}(x(t), y(t))$. Positive voltages locally switch the interface to a conducting state, while negative voltages locally restore the insulating state. Here, a conducting nanowire (shown in green) is being written. The conductance between the two electrodes is monitored by applying a small voltage bias on one of the two gold electrodes ($V_s$) and reading the current at the second electrode ($I_D$).

Figure 2. (Color online) Structure and electrical properties of non-reciprocal nanoscale devices at the LaAlO$_3$/SrTiO$_3$ interface. (a) Tip voltage profile during writing procedure. (b) Cross-sectional view illustrating the surface modulation-doping of the LaAlO$_3$/SrTiO$_3$ interface, resulting in a spatially uniform nanowire. (c) Schematic energy-band diagram for the conduction band minimum $E_c$ and Fermi energy $E_F$ for the uniform nanowire. (d) Current-voltage (*I-V*) plot for the uniform nanowire. (e) Tip voltage profile $V_{tip}$, (f) cross-sectional view, (g) schematic energy-band diagram and (h) *I-V* characteristics for a positive sawtooth potential $V_{tip}(x,y)= V_0 V_+(x,y)$, where $V_0$=10 V. (i) Tip voltage profile $V_{tip}$, (j) cross-sectional view, (k) schematic energy-band diagram and (l) *I-V* characteristics for a negative sawtooth potential $V_{tip}(x,y)= V_0 V_-(x,y)$, where $V_0$=10 V.

Figure 3. (Color online) (a) *I-V* plots for a nanowire cut at the same location multiple times with an AFM tip bias $V_{tip}$=-2 mV. Green curve indicates *I-V* curve before the first cut. Intermediate *I-V* curves are shown after every alternate cut. As the wire is cut, the potential barrier increases (inset), and the zero-bias conductance decreases; however, the overall *I-V* curve remains highly reciprocal. (b) *I-V* plots for a nanowire subject to a sequence of cuts $N_{cut}(x)$ at nine locations spaced 5 nm apart along the nanowire (see Table I). Green curve indicates *I-V* curve before the first cut. The asymmetry in $N_{cut}(x)$ results in a non-reciprocal *I-V* curve.

Table I. Number of cuts $N_{cut}(x)$ versus location *x*, resulting in non-reciprocal *I-V* profile in Fig. 3(b).

| $N_{cut}(x)$ | 1 | 3 | 5 | 7 | 11 | 15 | 21 | 31 | 38 |
|---|---|---|---|---|---|---|---|---|---|
| Position *x* (nm) | 0 | 5 | 10 | 15 | 20 | 25 | 30 | 35 | 40 |

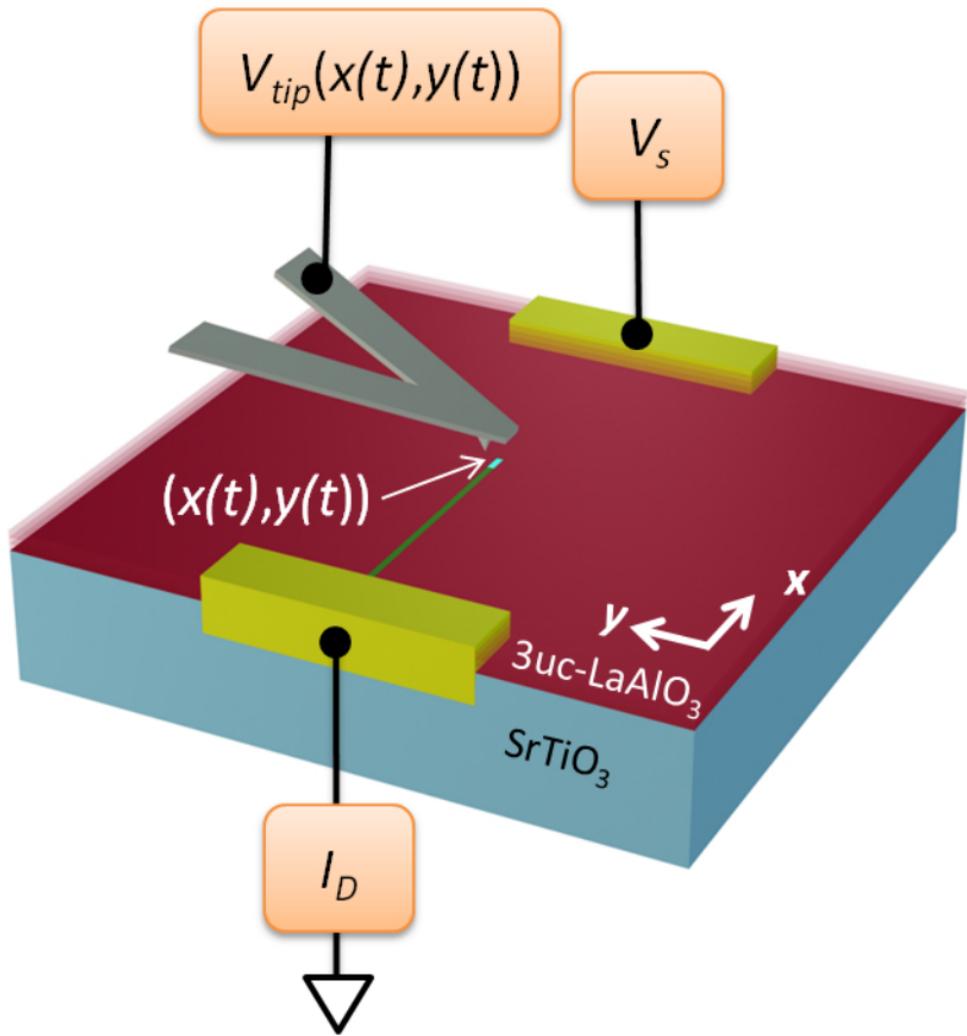

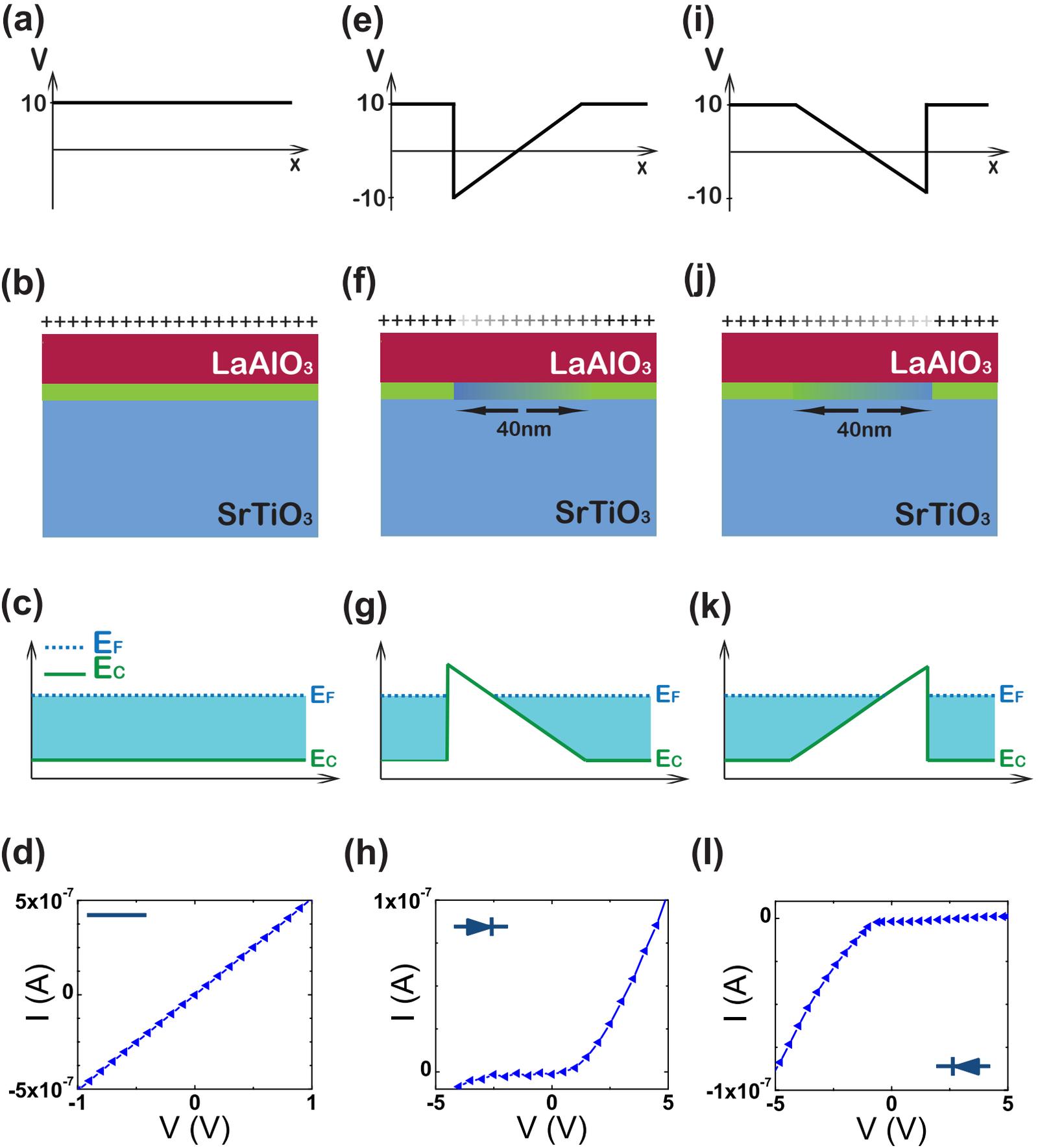

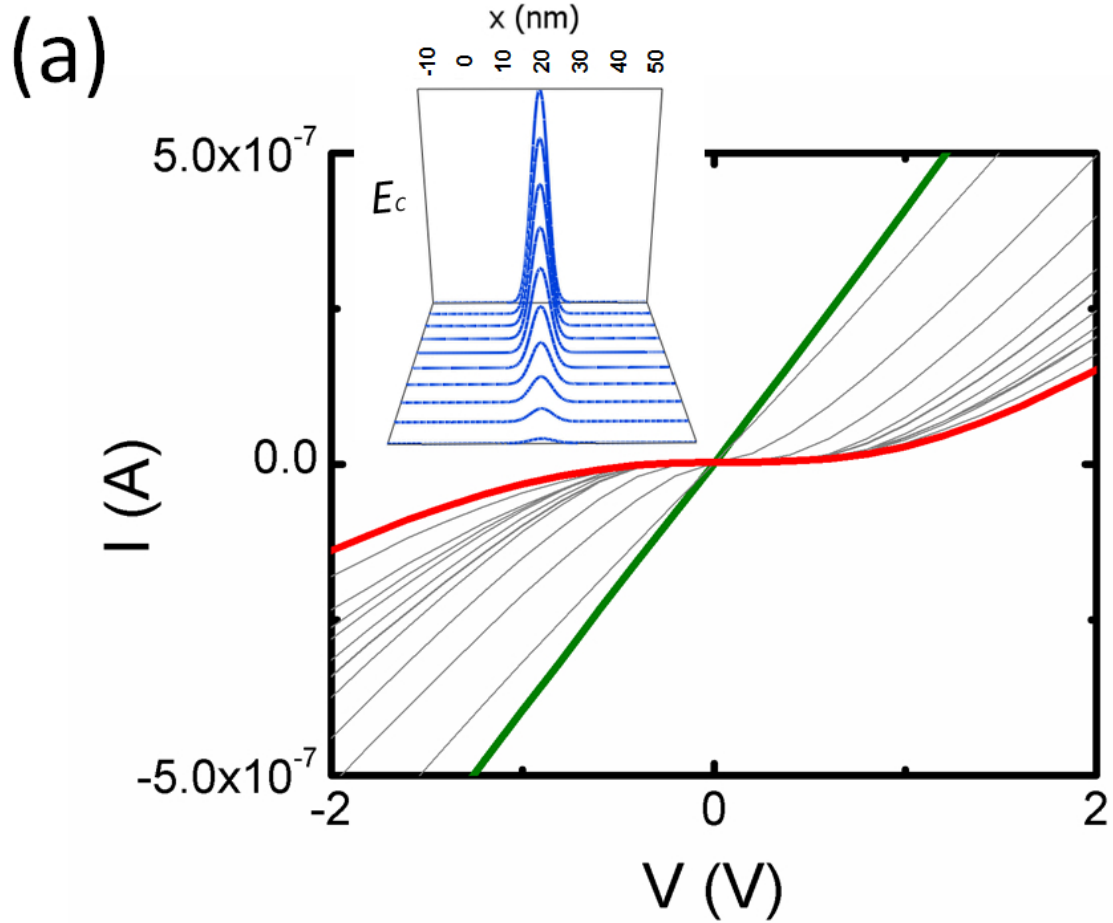

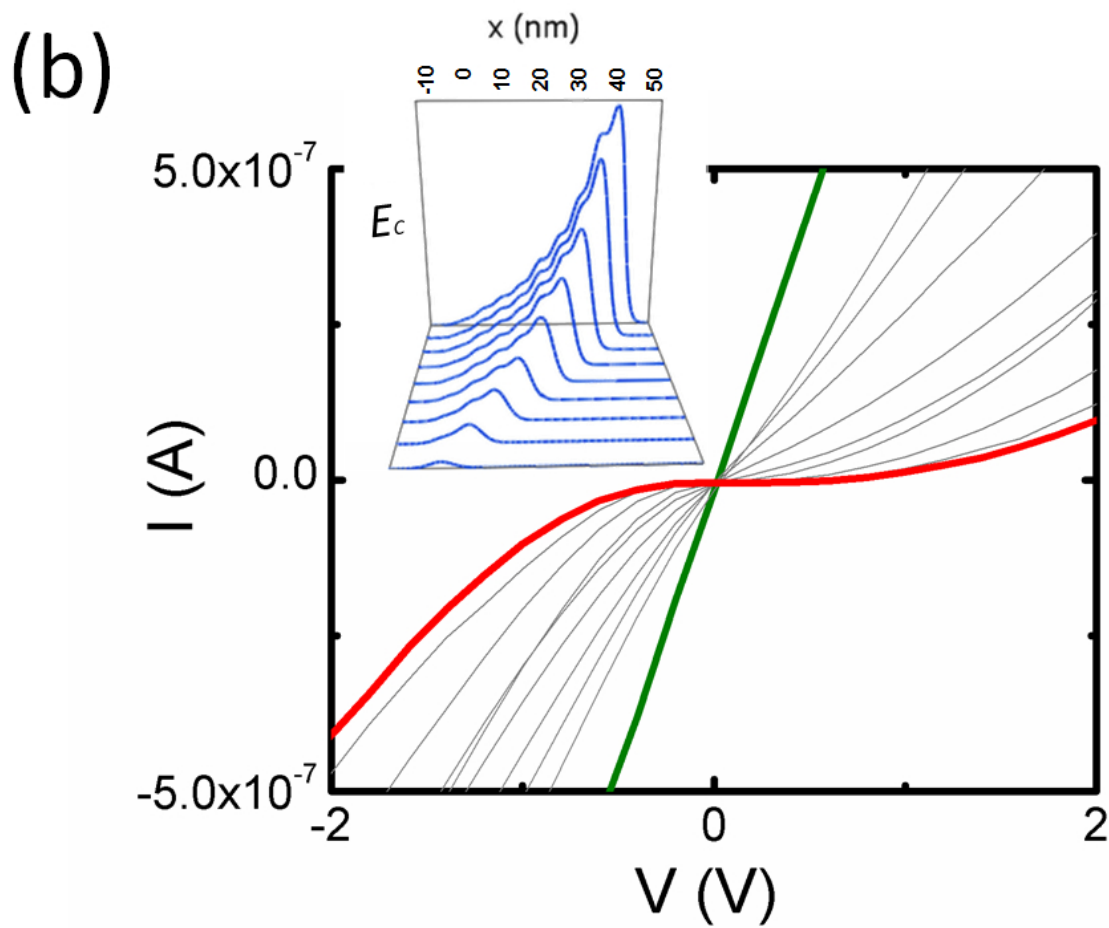